\documentclass[aps,prl,superscriptaddress,twocolumn,nofootinbib,floatfix]{revtex4} 
\usepackage{graphicx}

\begin{document}
\title{Searching For New Physics With \boldmath$B\to K\pi$ Decays}

\author{M.~Ciuchini}
\affiliation{INFN, Sezione di Roma Tre, I-00146 Roma, Italy}
\author{E.~Franco}
\affiliation{INFN, Sezione di Roma, I-00185 Roma, Italy}
\author{G.~Martinelli}
\affiliation{INFN, Sezione di Roma, I-00185 Roma, Italy}
\affiliation{Dipartimento di Fisica, Universit\`a di Roma ``La
  Sapienza'', I-00185 Roma, Italy} 
\author{M.~Pierini}
\affiliation{CERN, CH-1211 Geneva 23, Switzerland}
\author{L.~Silvestrini}
\affiliation{INFN, Sezione di Roma, I-00185 Roma, Italy}

\begin{abstract}
  We propose a method to quantify the Standard Model uncertainty in
  $B \to K \pi$ decays using the experimental data,
  assuming that power counting provides a reasonable estimate of the
  subleading terms in the $1/m_b$ expansion. Using this method, we
  show that present $B \to K \pi$ data are compatible with the
  Standard Model. We analyze the pattern of subleading terms required
  to reproduce the $B\to K\pi$ data and argue that anomalously large
  subleading terms are not needed. Finally, we find that $S_{K_S\pi^0}$
  is fairly insensitive to hadronic uncertainties and obtain the Standard
  Model estimate $S_{K_S\pi^0}=0.74\pm 0.04$.
\end{abstract}

\maketitle

A decade of physics studies at the $B$ factories produced the
impressive set of results on $B\to K\pi$ decays summarized in
Table~\ref{tab:res}. As data became more and more accurate,
phenomenological analyses based on flavour symmetries and/or hadronic
models were not able to fully reproduce the data. This led several
authors to introduce the $K\pi$ \textit{puzzle} in its different
incarnations~\cite{puzzle,puzzle2}. In particular, the difference
$\Delta A_\mathrm{CP}=A_\mathrm{CP}(K^+\pi^0)-A_\mathrm{CP}(K^+\pi^-)$
has recently received considerable attention, following the new
measurement $\Delta A_\mathrm{CP}=0.164\pm 0.037$ published by the
Belle collaboration~\cite{NATUA.452.332}. It has been argued that
$\Delta A_\mathrm{CP}$ could be a hint of New Physics (NP), but
alternative explanations within the Standard Model (SM) have also been
considered.

\begin{table}
\begin{center}
\begin{tabular}{|c|c|c|c|}
\hline
Decay Mode & HFAG average & global fit & fit prediction\\\hline
$10^6$ BR$(K^+ \pi^-)$ & $19.4 \pm 0.6$ & $19.5   \pm 0.5$   & $19.7
\pm 1.0$   \\ 
$10^6$ BR$(K^+ \pi^0)$ & $12.9 \pm 0.6$ & $12.7   \pm 0.5$   & $12.4
\pm 0.7$   \\ 
$10^6$ BR($K^0 \pi^+)$ & $23.1 \pm 1.0$ & $23.8   \pm 0.8$   & $24.9
\pm 1.2$   \\ 
$10^6$ BR($K^0 \pi^0)$ & $9.8 \pm 0.6$ & $9.3    \pm 0.4$   & $8.7
\pm 0.6$   \\ 
$A_\mathrm{CP}(K^+ \pi^-)$ [\%] & $-9.8 \pm 1.2$ & $-9.5 \pm 1.2$ &
$3.9 \pm 6.8$  \\ 
$A_\mathrm{CP}(K^+ \pi^0)$ [\%] & $5.0  \pm 2.5$ & $3.6  \pm 2.4$ &
$-6.2 \pm 6.0$ \\ 
$A_\mathrm{CP}(K^0 \pi^+)$ [\%] & $0.9  \pm 2.5$ & $1.8  \pm 2.1$ & $
6.2 \pm 4.5$ \\ 
$C(K_S \pi^0)$    & $0.01  \pm 0.10$ & $0.09   \pm 0.03$  & $ 0.10 \pm
0.03$ \\ 
$S(K_S \pi^0)$   & $0.57 \pm 0.17$ & $0.73   \pm 0.04$  & $ 0.74 \pm
0.04$ \\
\hline
$\Delta A_\mathrm{CP}$ [\%] & $ 14.8\pm 2.8$ & $13.1  \pm 2.6$ & $1.7
\pm 6.1$\\ 
\hline
\end{tabular}
\caption{Experimental inputs and fit results for $B\to K\pi$. For each
  observable, we report experimental results (BR$^\mathrm{exp}$ and
  $A_\mathrm{CP}^\mathrm{exp}$)~\cite{sk0p0,NATUA.452.332,brexp}
  taken from HFAG~\cite{hfag}, the results of the fit using all the
  constraints (third column) and the prediction obtained using all
  constraints except the considered observable (fourth column). For
  $\Delta A_\mathrm{CP}$, the prediction is obtained by removing both
  $A_\mathrm{CP}(K^+ \pi^0)$ and
  $A_\mathrm{CP}(K^+ \pi^-)$ from the fit.}
\label{tab:res}
\end{center}
\end{table}

To understand whether $B\to K\pi$ decays are really puzzling, possibly
calling for NP, one has to control the SM expectations for the $B\to
K\pi$ amplitudes with a level of accuracy dictated by the size of the
potential NP contributions.  Thanks to the progress of theory in the
last few years, we know that two-body non-leptonic $B$ decay
amplitudes are factorizable in the infinite $b$-quark mass limit, {\em
  i.e.}\ computable in terms of a reduced set of universal
non-perturbative parameters~\cite{qcdf,pqcd,scet}. However, the
accuracy of the predictions obtained with factorization is limited by
the uncertainties on the non-perturbative parameters on the one hand
and by the uncalculable subleading terms in the $1/m_b$ expansion on
the other. The latter problem is particulary severe for $B\to K\pi$
decays where some power-suppressed terms are doubly Cabibbo-enhanced
with respect to factorizable terms~\cite{charming}. Indeed
factorization typically predicts too small $B \to K \pi$ branching
ratios, albeit with large uncertainties. The introduction of
subleading terms, certainly present at the physical value of the $b$
quark mass, produces large effects in branching ratios and CP
asymmetries, leading to a substantial model dependence of the SM
predictions.  Given this situation, NP contributions to $B\to K\pi$
amplitudes could be easily misidentified.

In this paper, we suggest a method to estimate the SM
uncertainty given the experimental data, assuming that subleading
terms are at most of order $1/m_b$.~\footnote{An early attempt
at this method was presented in ref.~\cite{pieriniCKM06}.}
This procedure provides a
solid starting point for NP searches. Clearly, we are not sensitive
to the presence of NP contributions of the same size as the
subleading corrections to factorization. 

We now describe our method in detail. We start with a general
parametrization of the $B\to K\pi$ amplitudes derived from the one in
ref.~\cite{buras}. The decay amplitudes are given by:
\begin{eqnarray}
    A(B^+\to K^0\pi^+) &=& -V_{ts}V_{tb}^* P + V_{us} V_{ub}^*
    A\,,\nonumber \\
    A(B^+\to K^+\pi^0) &=& \frac{1}{\sqrt{2}} \bigl( V_{ts}V_{tb}^*
      (P+\Delta P_1+\Delta P_2) 
      -\nonumber \\
      &&\qquad V_{us} V_{ub}^* (E_1+E_2+A)\bigr)\,,\nonumber\\
    A(B^0\to K^+\pi^-) &=& V_{ts}V_{tb}^* (P+\Delta P_1)-V_{us}
    V_{ub}^* E_1\,,\nonumber\\ 
    A(B^0\to K^0\pi^0) &=& -\frac{1}{\sqrt{2}} \bigl( V_{ts}V_{tb}^*
      (P-\Delta P_2)+\nonumber \\
      &&\qquad \quad V_{us} V_{ub}^* E_2\bigr)\,.
      \label{eq:ampli}
\end{eqnarray}
In terms of the parameters of ref.~\cite{buras}, our parameters read
\begin{eqnarray}
    E_1 &=& E_1(s, q, q; B, K, \pi)-P_1^\mathrm{GIM}(s, q; B, K,
    \pi)\,,\nonumber\\
    E_2 &=& E_2(q, q, s; B, \pi, K)+P_1^\mathrm{GIM}(s, q; B, K,
    \pi)\,,\nonumber\\ 
    A   &=& A_1(s, q, q; B, K, \pi)-P_1^\mathrm{GIM}(s, q; B, K,
    \pi)\,,\nonumber\\ 
    P   &=& P_1(s, d; B, K, \pi)\,,\nonumber\\
    \Delta P_1 &=& P_1(s, u; B, K, \pi)-P_1(s, d; B, K,
    \pi)\,,\nonumber\\ 
    \Delta P_2 &=& P_2(s, u; B, \pi , K)-P_2(s, d; B, \pi, K)\,.
    \label{eq:params}
\end{eqnarray}
With respect to the most general parametrization, we have neglected
isospin breaking in the hadronic matrix elements of the effective weak
Hamiltonian, yet fully retaining the effects of the electroweak penguins
(EWP). This assumption reduces the number of independent parameters
and removes the dependence on meson charges in the arguments of the
parameters on the {r.h.s.} of eqs.~(\ref{eq:params}), where $q$
denotes the light quarks.

\begin{figure}[tb]
 \begin{center}
 \includegraphics[width=.23\textwidth]{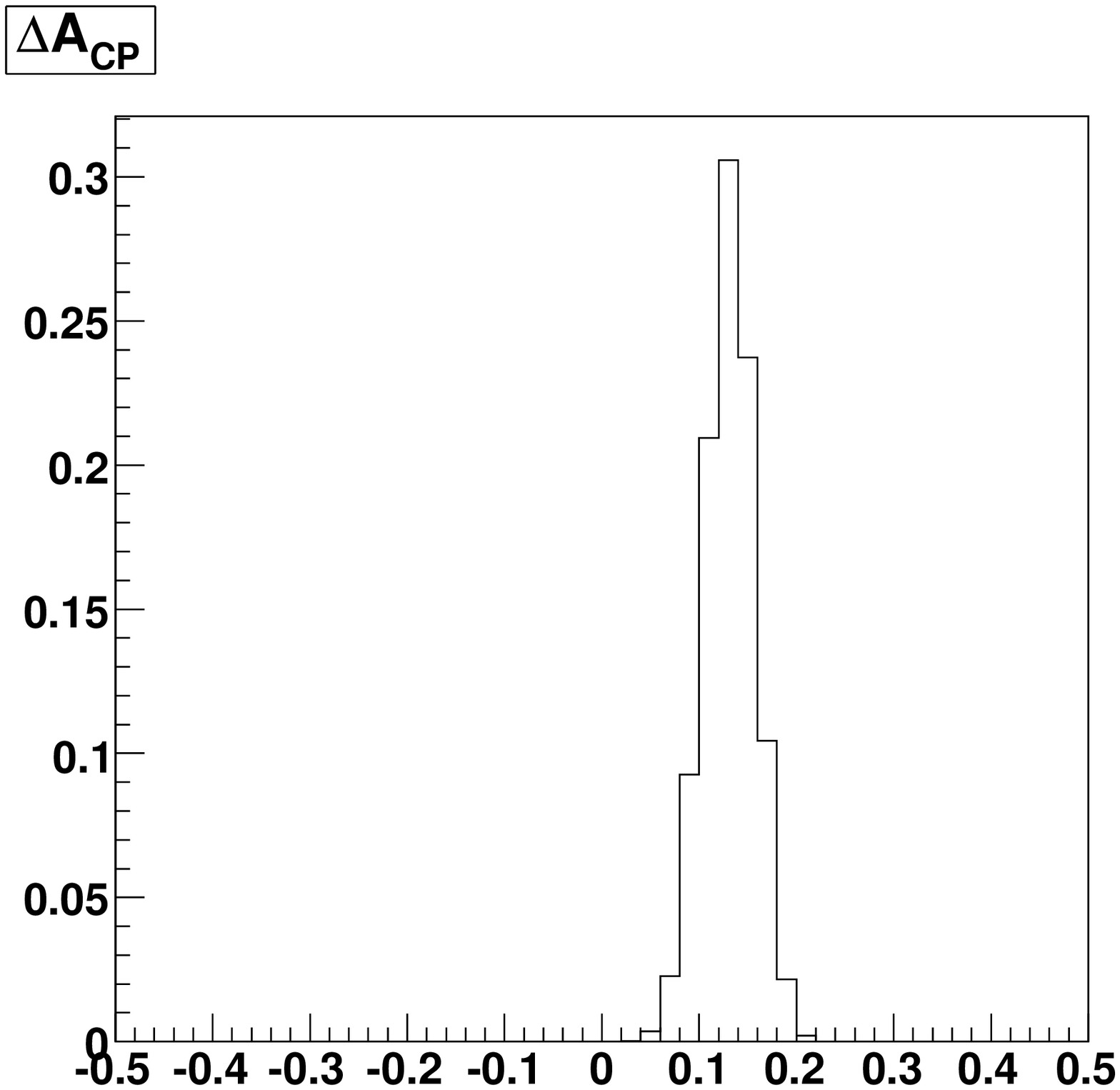}
 \includegraphics[width=.23\textwidth]{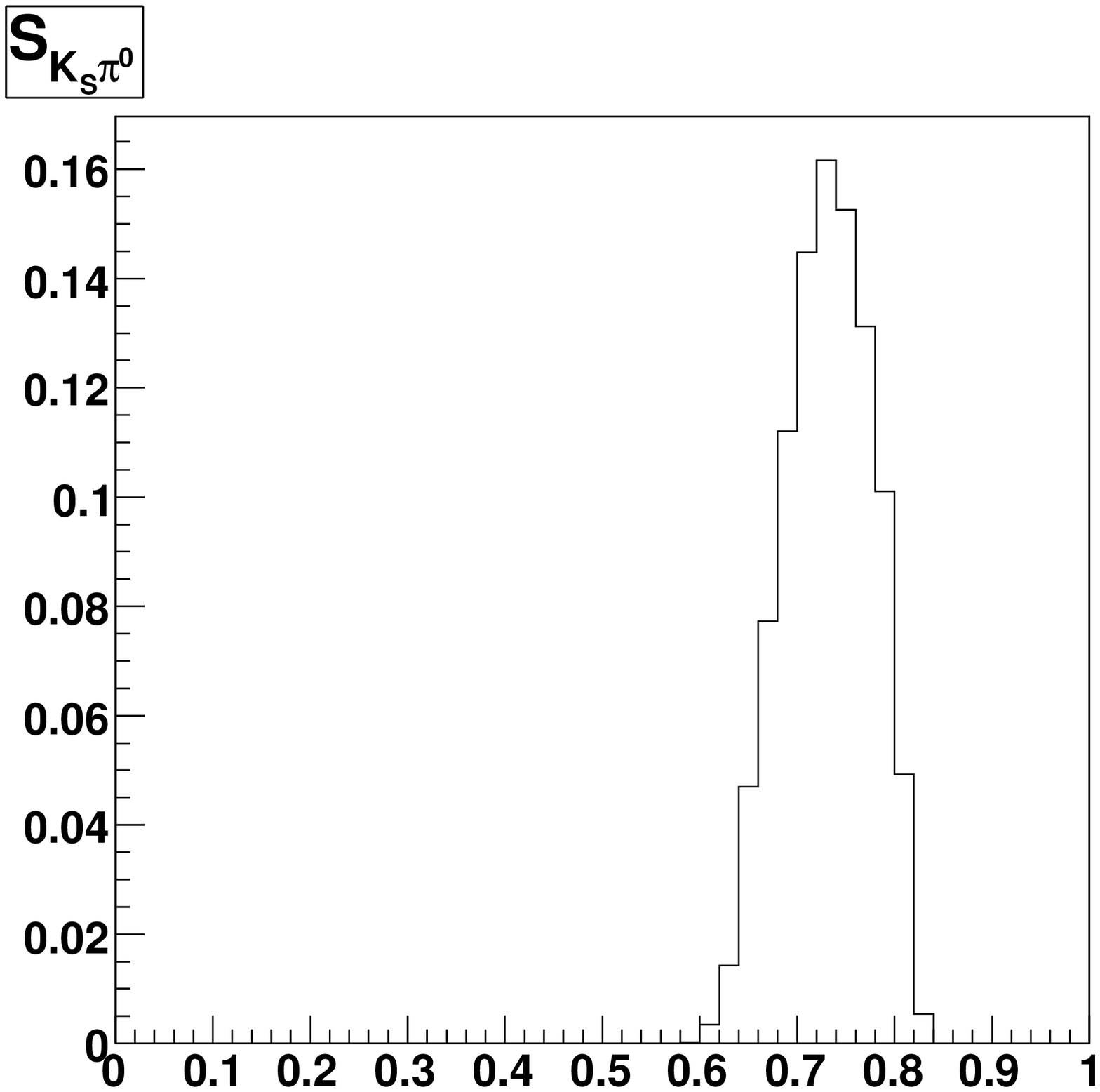}
\end{center}
\caption{P.d.f. obtained from the global  fit for $\Delta A_\mathrm{CP}$ (left) and for $S(K_S \pi^0)$ (right).}
\label{fig:acpdiff_sk0pi0}
\end{figure}

\begin{figure}[b!]
 \begin{center}
 \includegraphics[width=.32\textwidth]{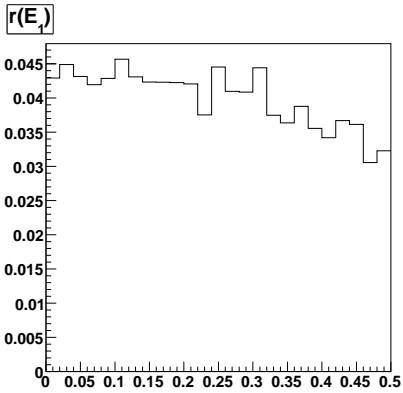}
  \end{center}
\caption{P.d.f. obtained from the global fit for the parameter
  $r(E_1)$ defined in 
  eqs.~(\ref{eq:params1}).\label{fig:had0}}
\end{figure}

\begin{figure*}[tb]
 \begin{center}
 \includegraphics[width=.32\textwidth]{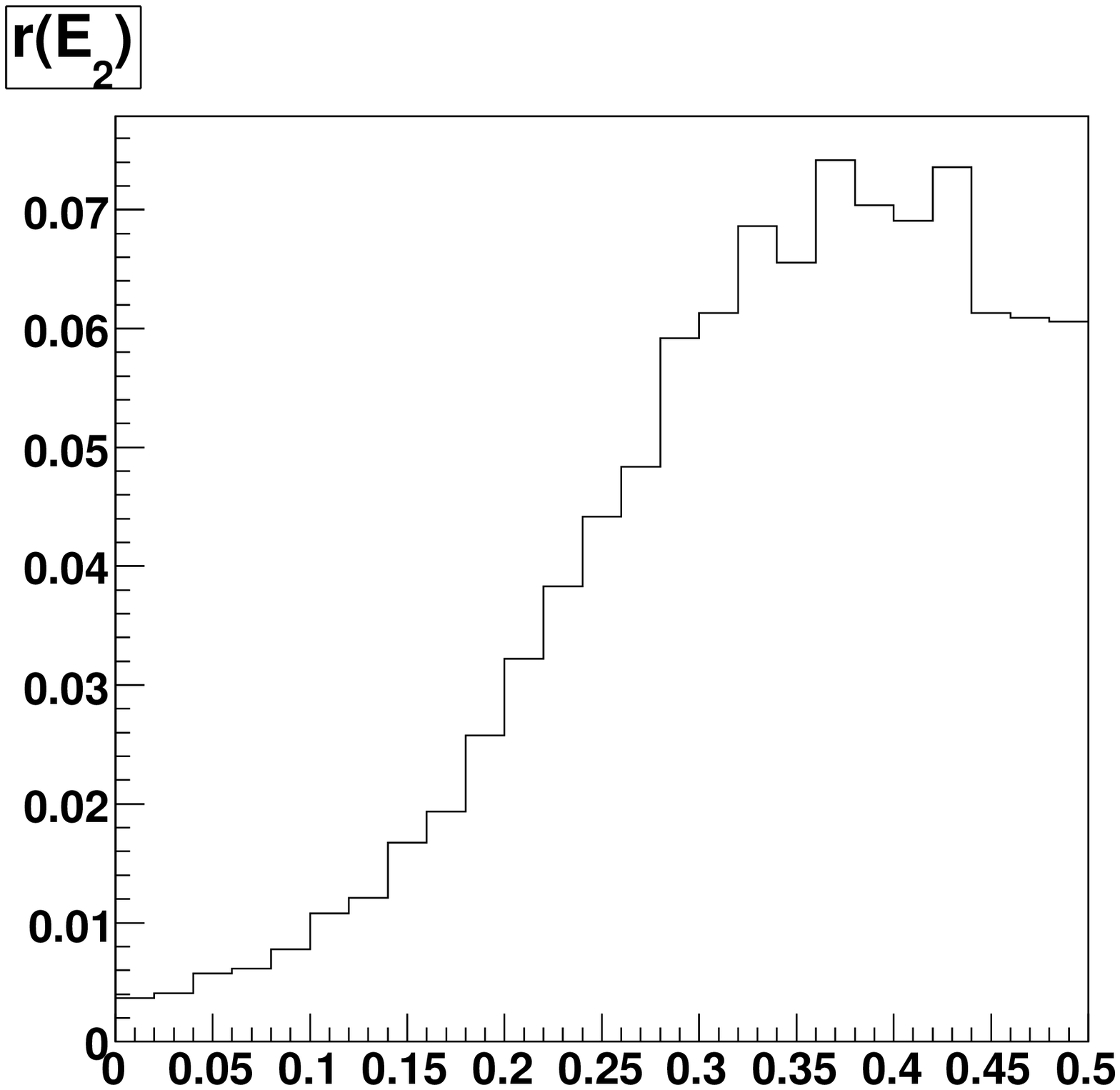}
 \includegraphics[width=.32\textwidth]{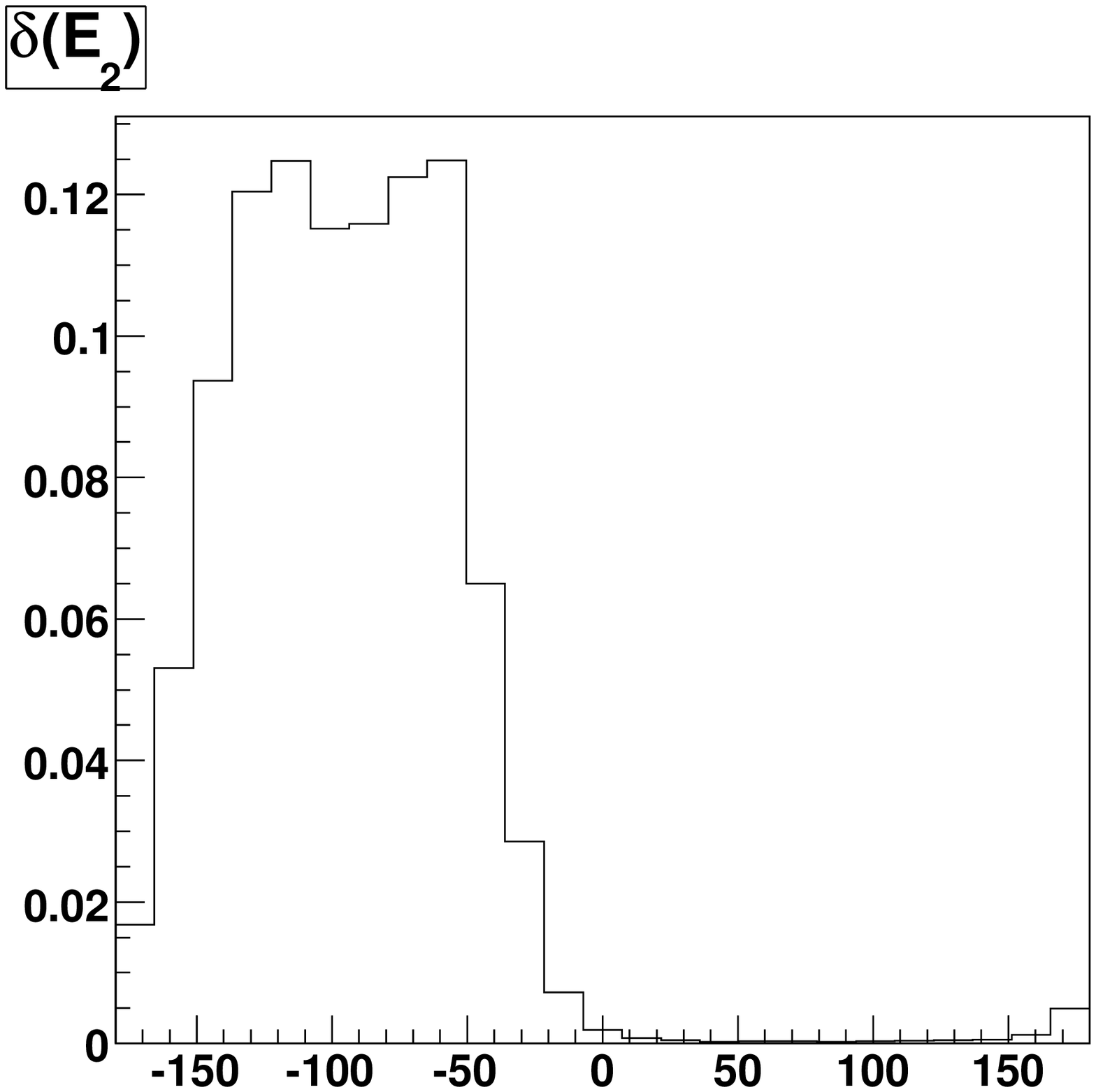}
 \includegraphics[width=.32\textwidth]{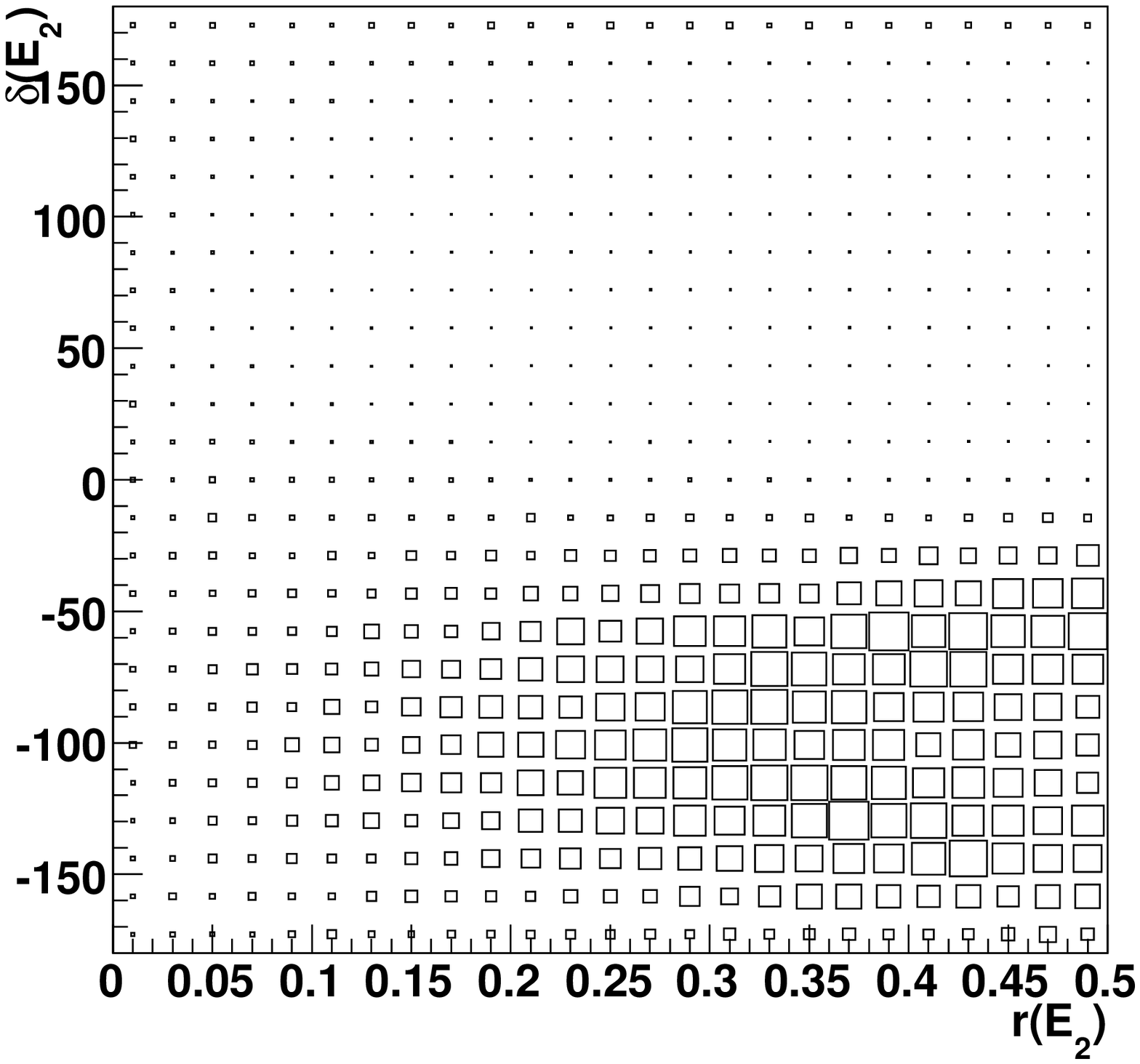}
 \includegraphics[width=.32\textwidth]{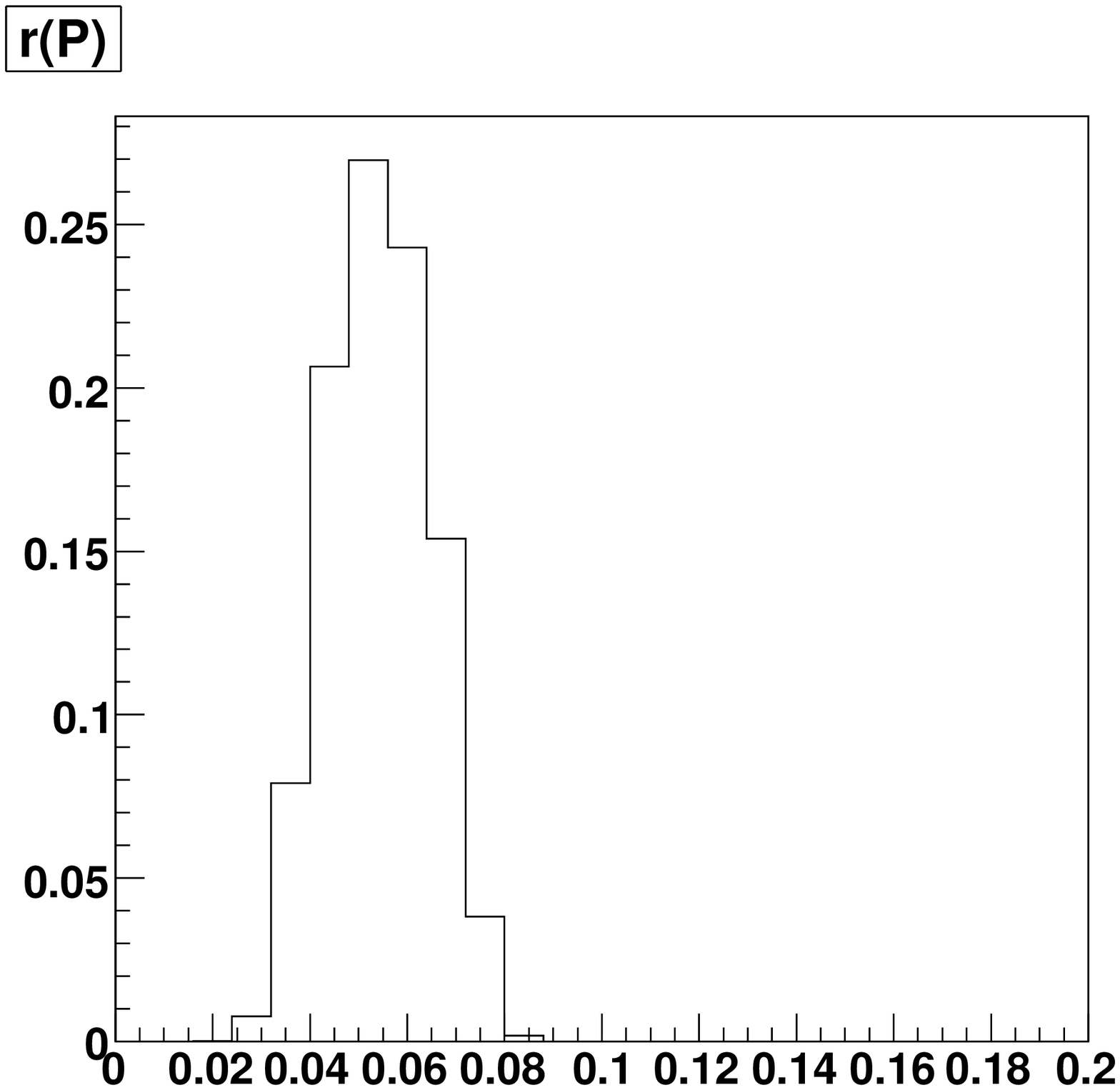}
 \includegraphics[width=.32\textwidth]{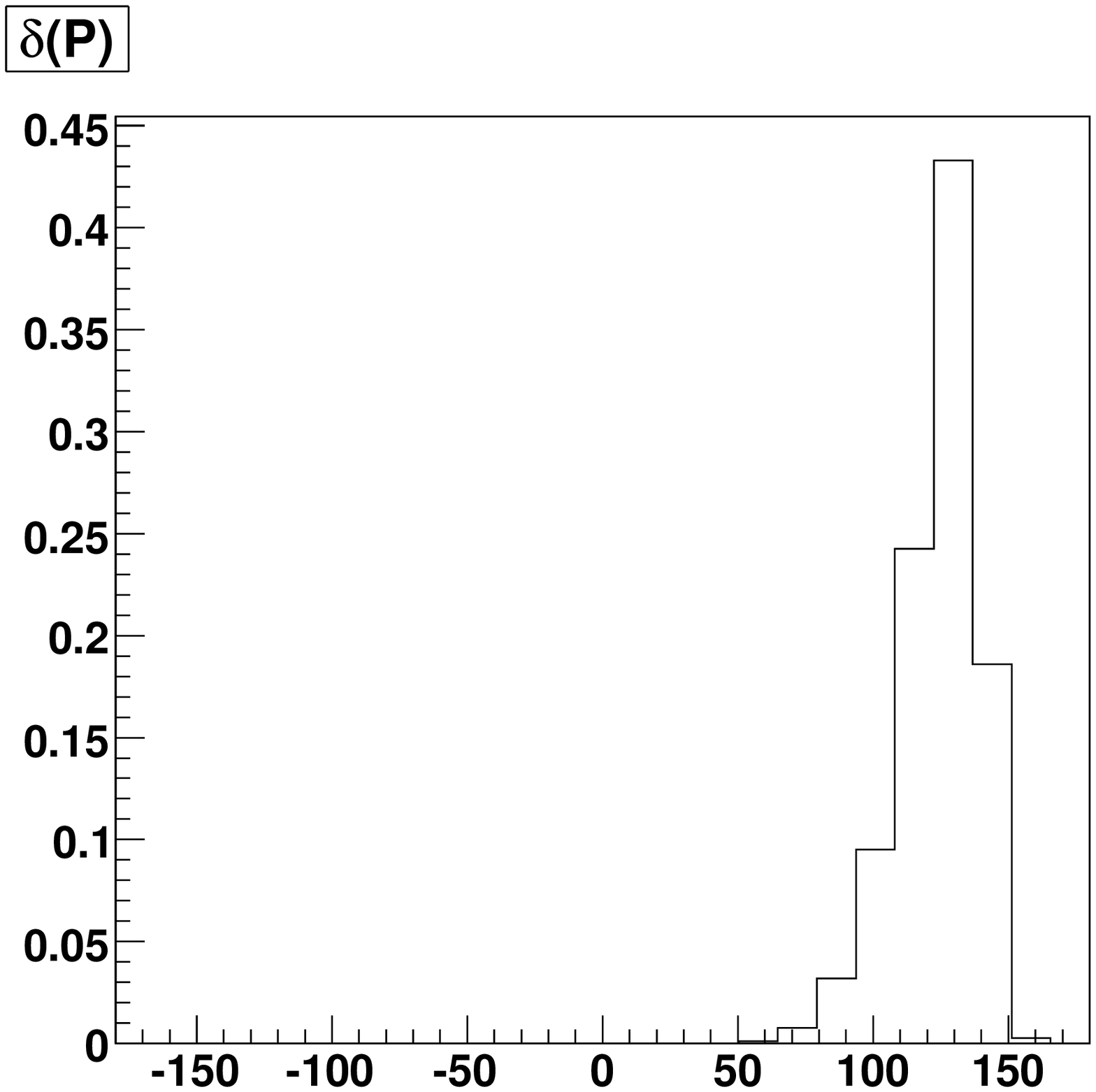}
 \includegraphics[width=.32\textwidth]{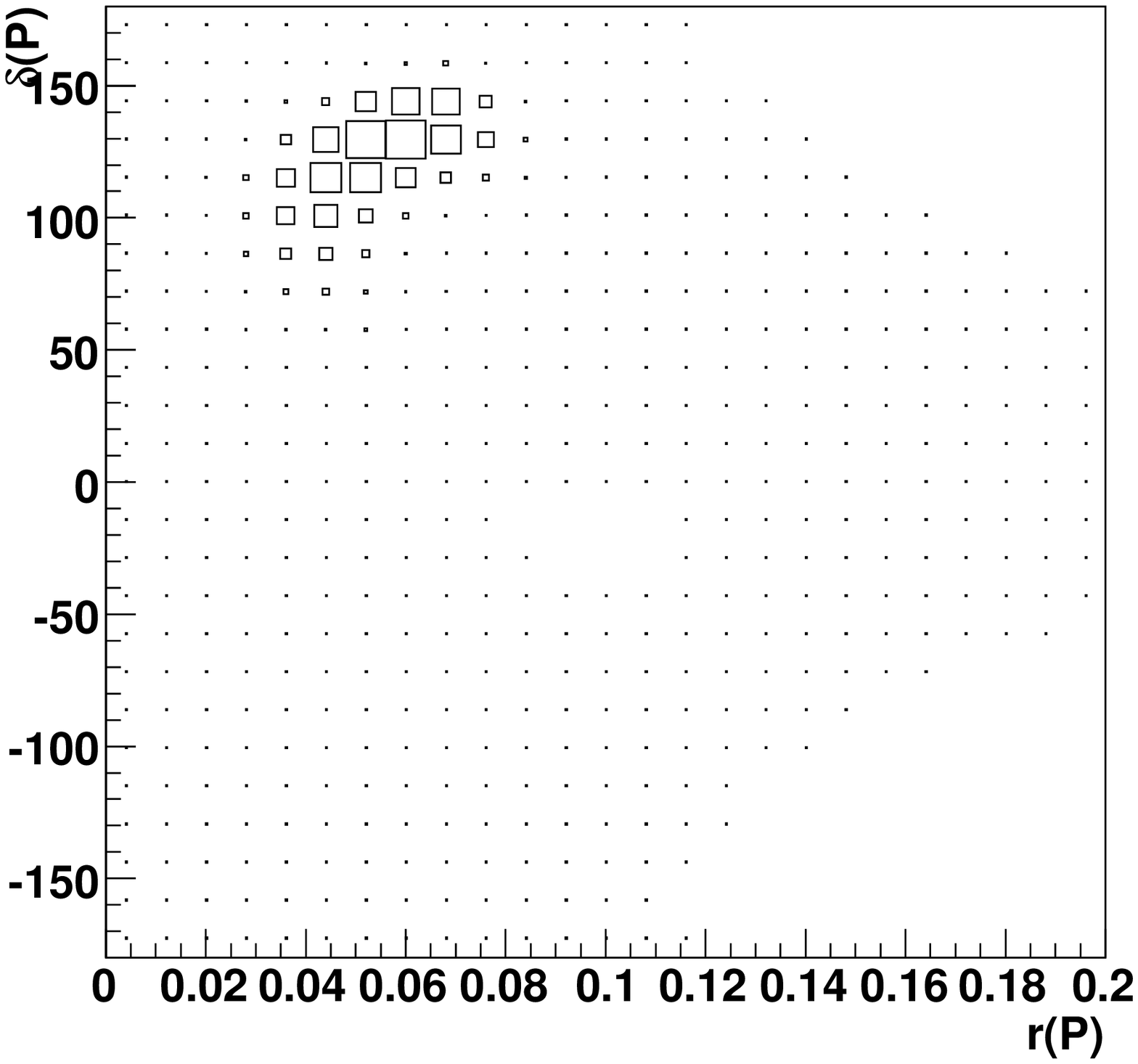}
 \includegraphics[width=.32\textwidth]{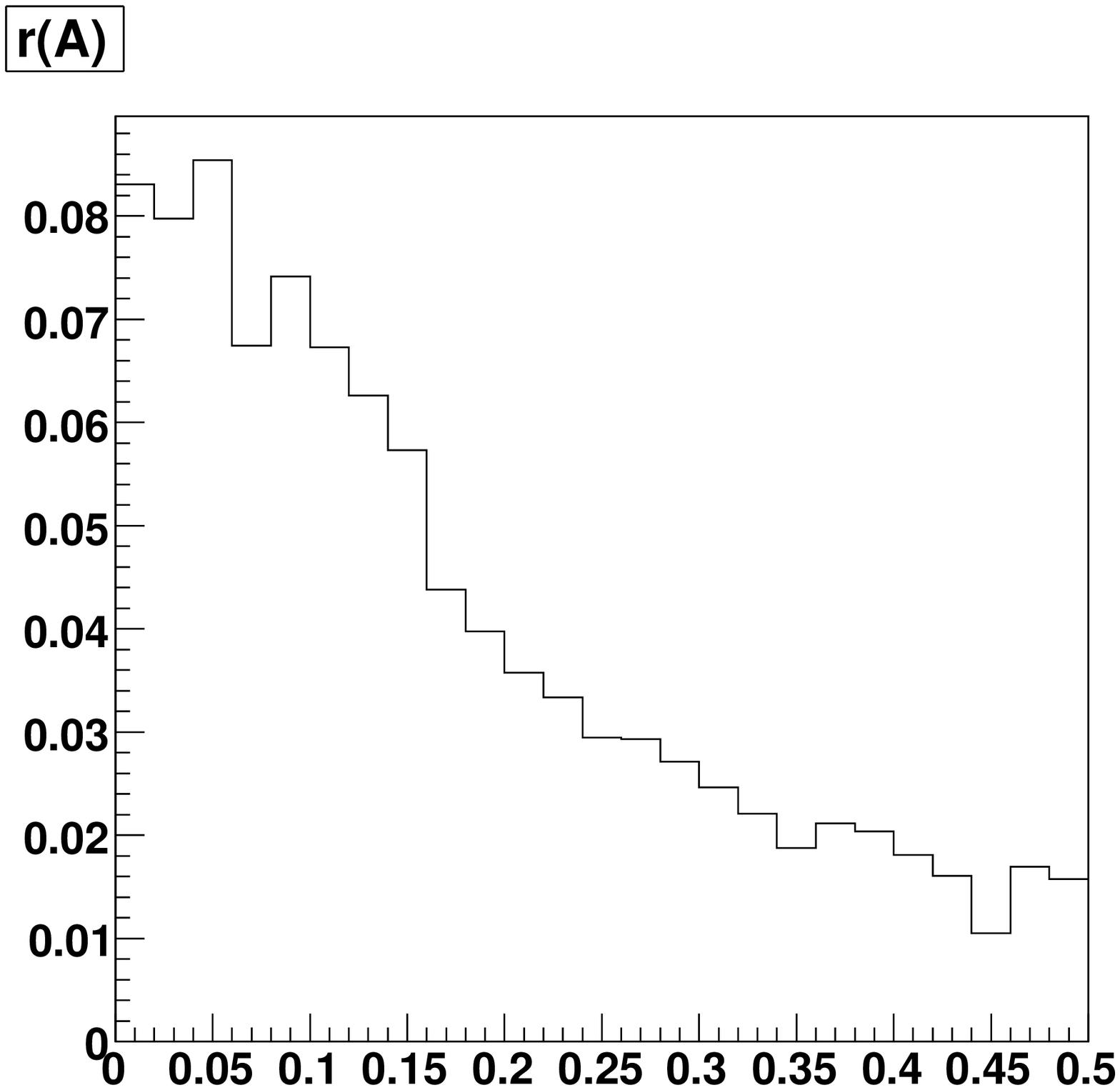}
 \includegraphics[width=.32\textwidth]{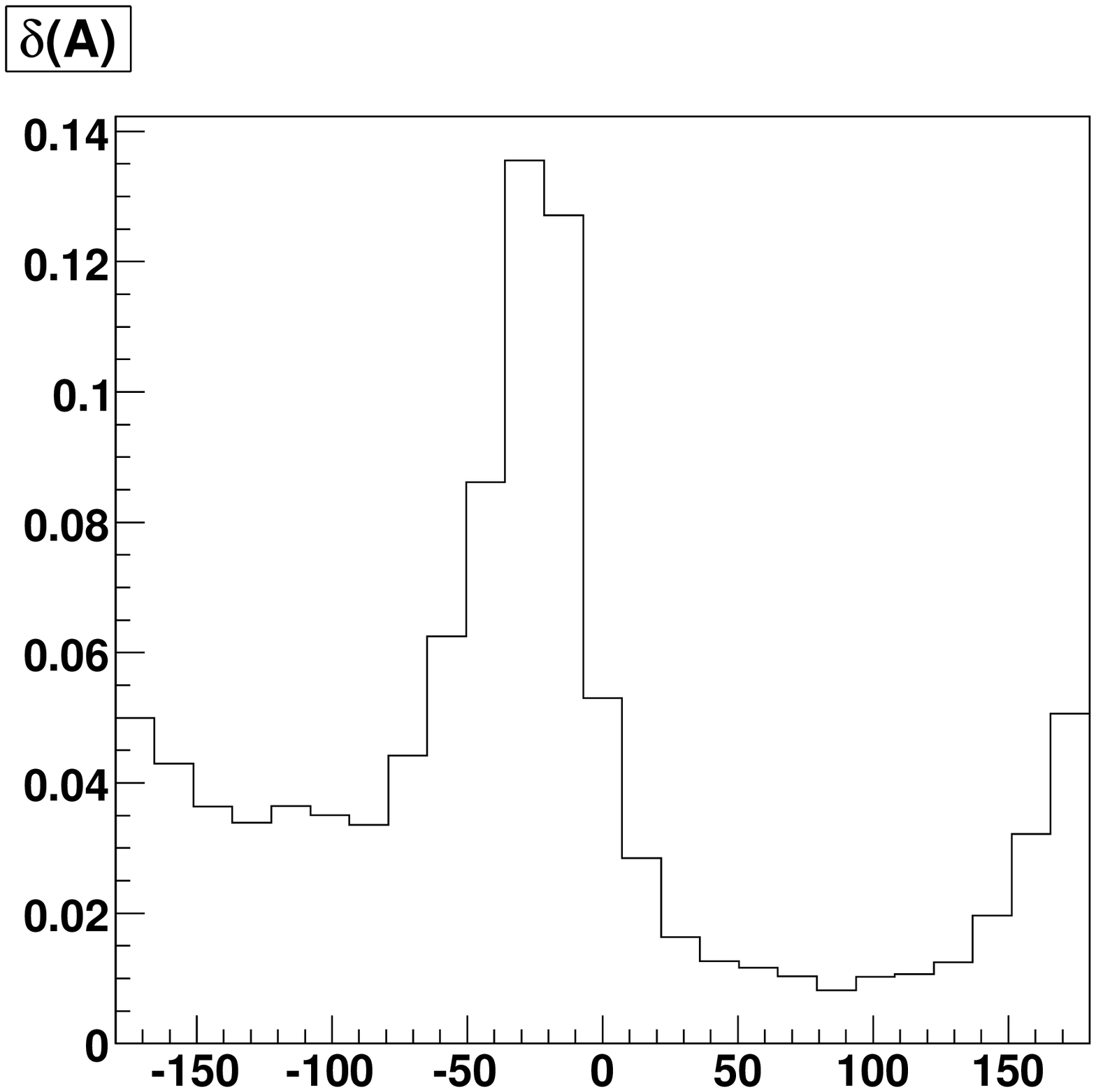}
 \includegraphics[width=.32\textwidth]{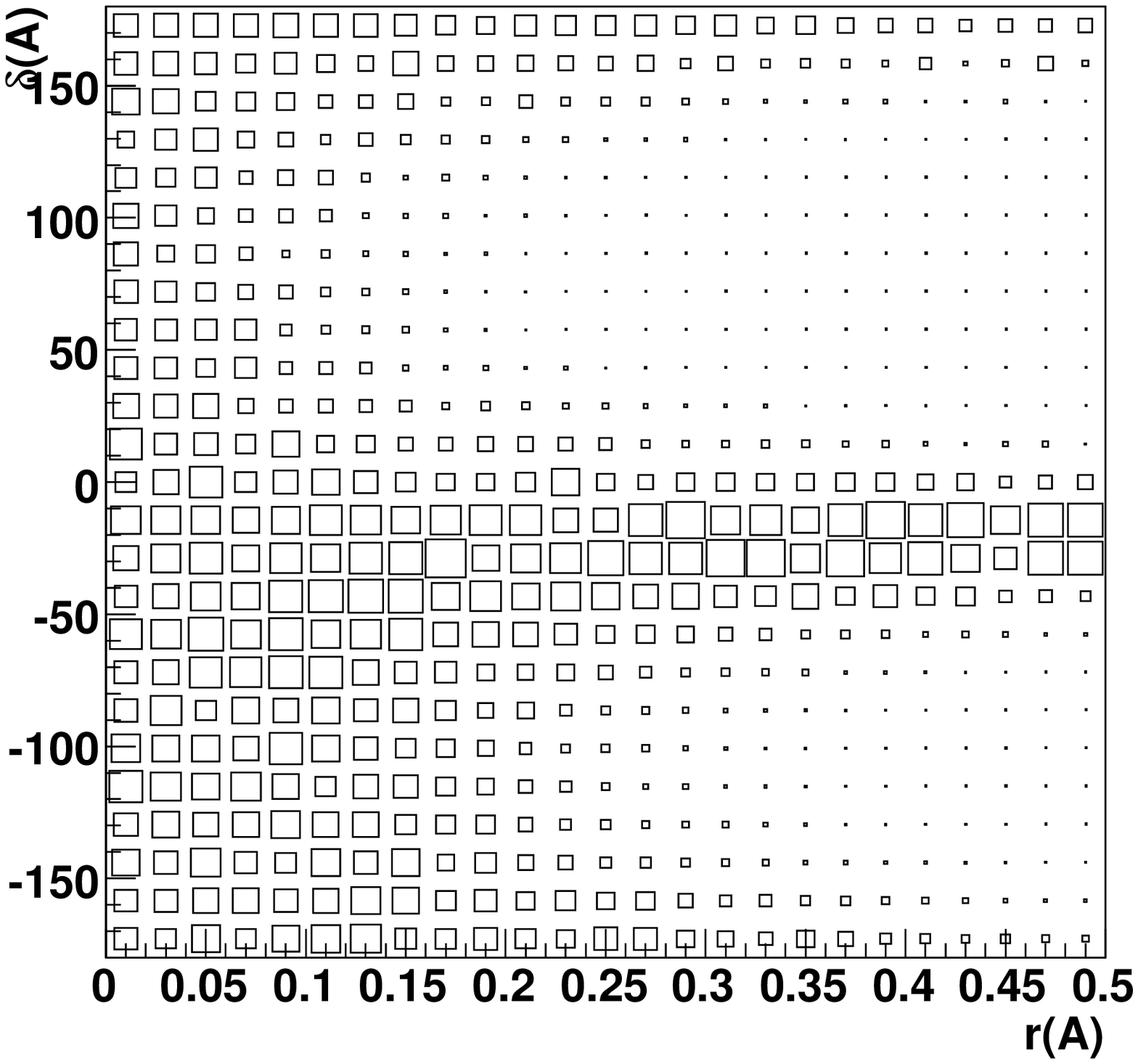}
 \end{center}
\caption{1D and 2D p.d.f.'s obtained from the global fit for the
  parameters $r(E_2)$, $\delta(E_2)$, $r(P)$, $\delta(P)$, and $r(A)$,
  $\delta(A)$ defined in 
  eqs.~(\ref{eq:params1}).\label{fig:had1}}
\end{figure*}

\begin{figure*}[tb]
 \begin{center}
 \includegraphics[width=.32\textwidth]{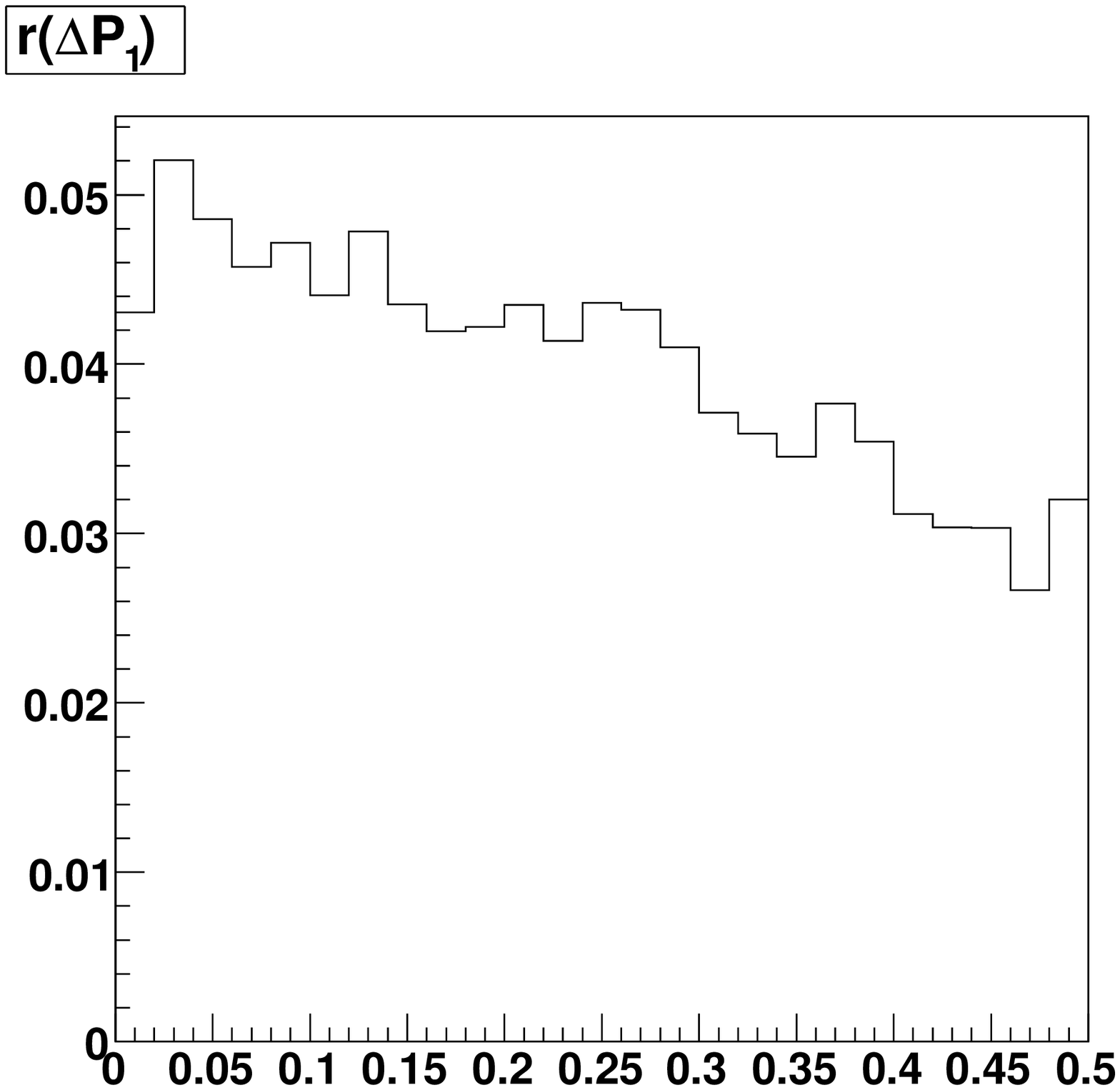}
 \includegraphics[width=.32\textwidth]{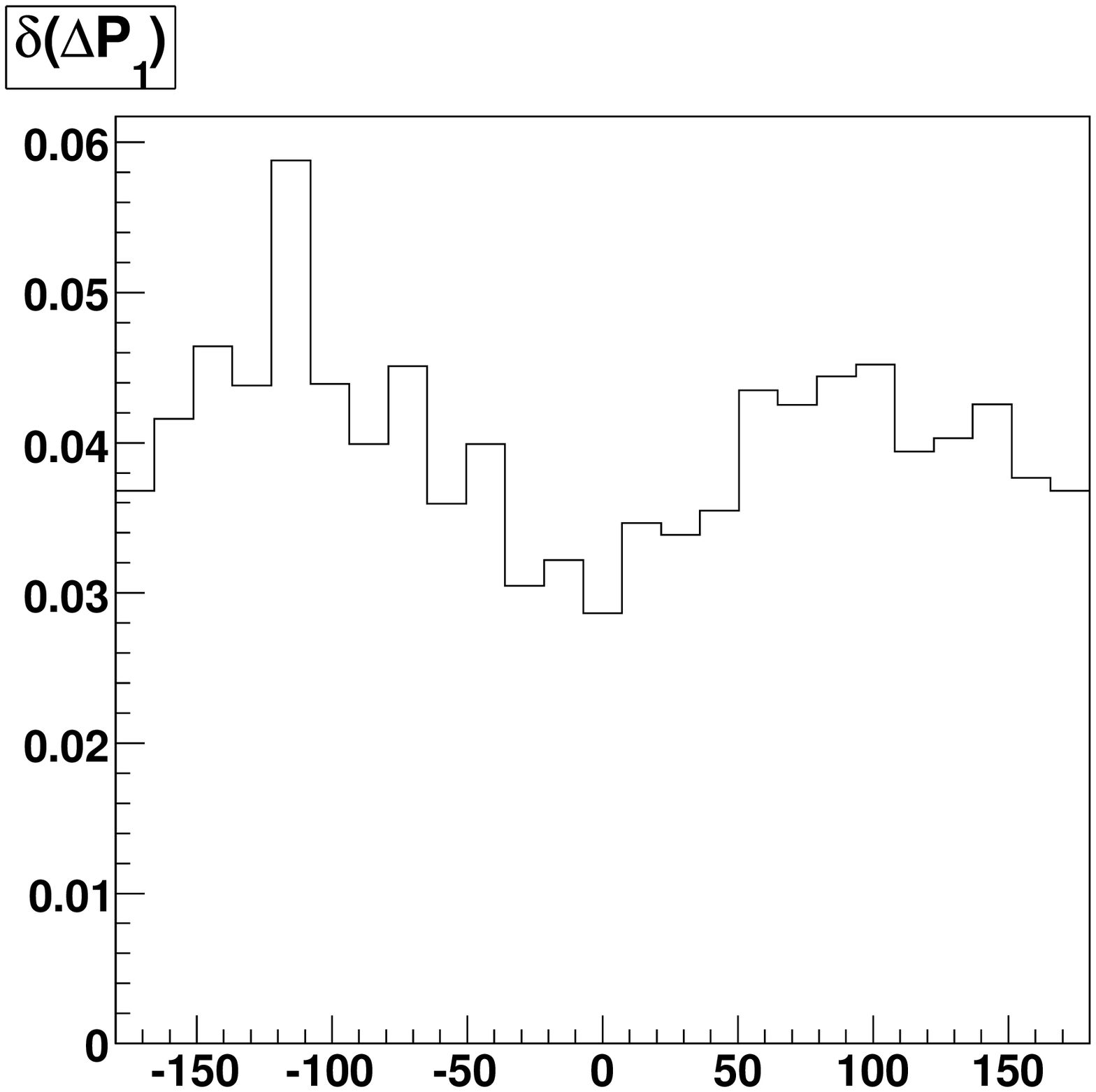}
 \includegraphics[width=.32\textwidth]{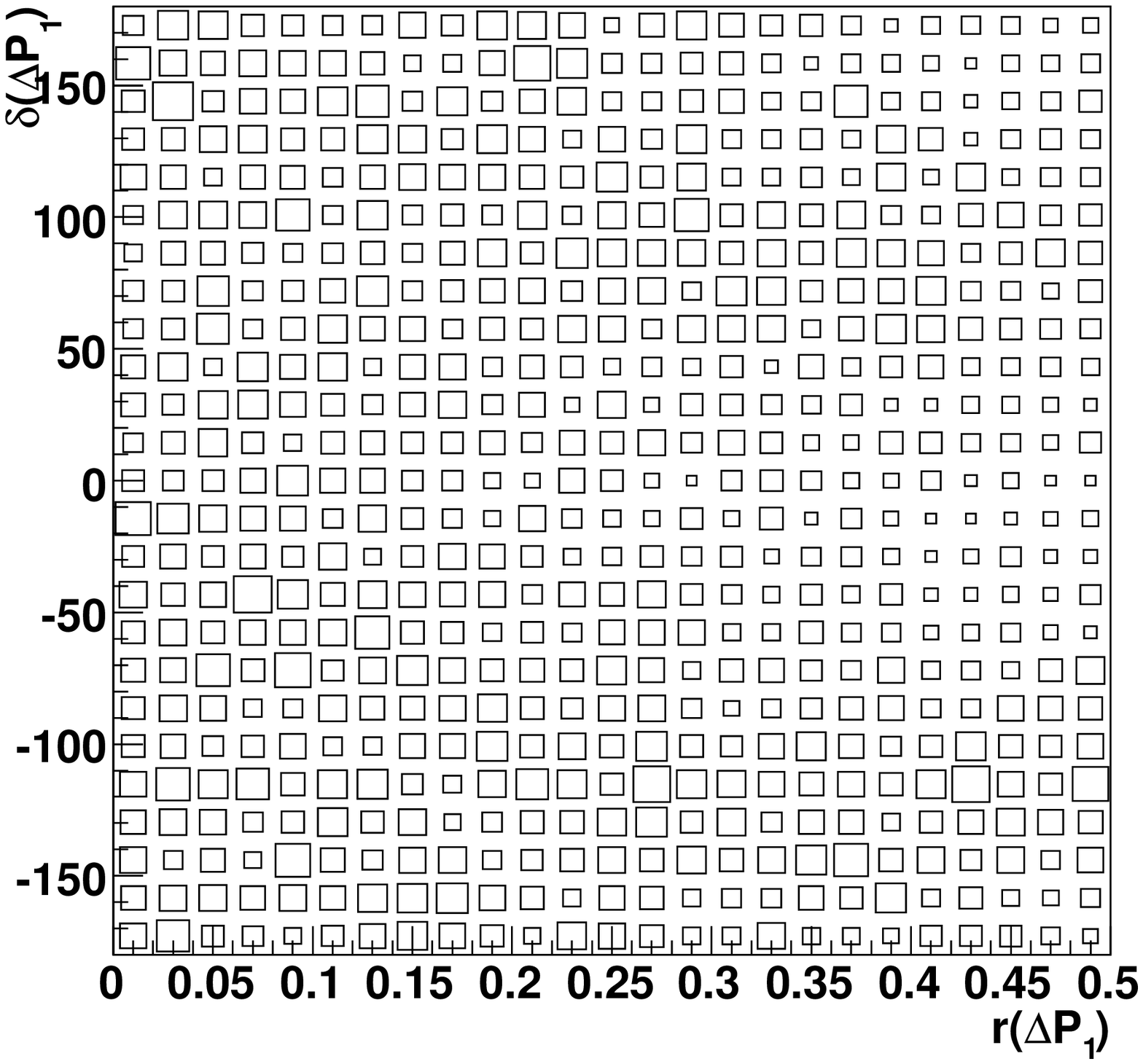}
 \includegraphics[width=.32\textwidth]{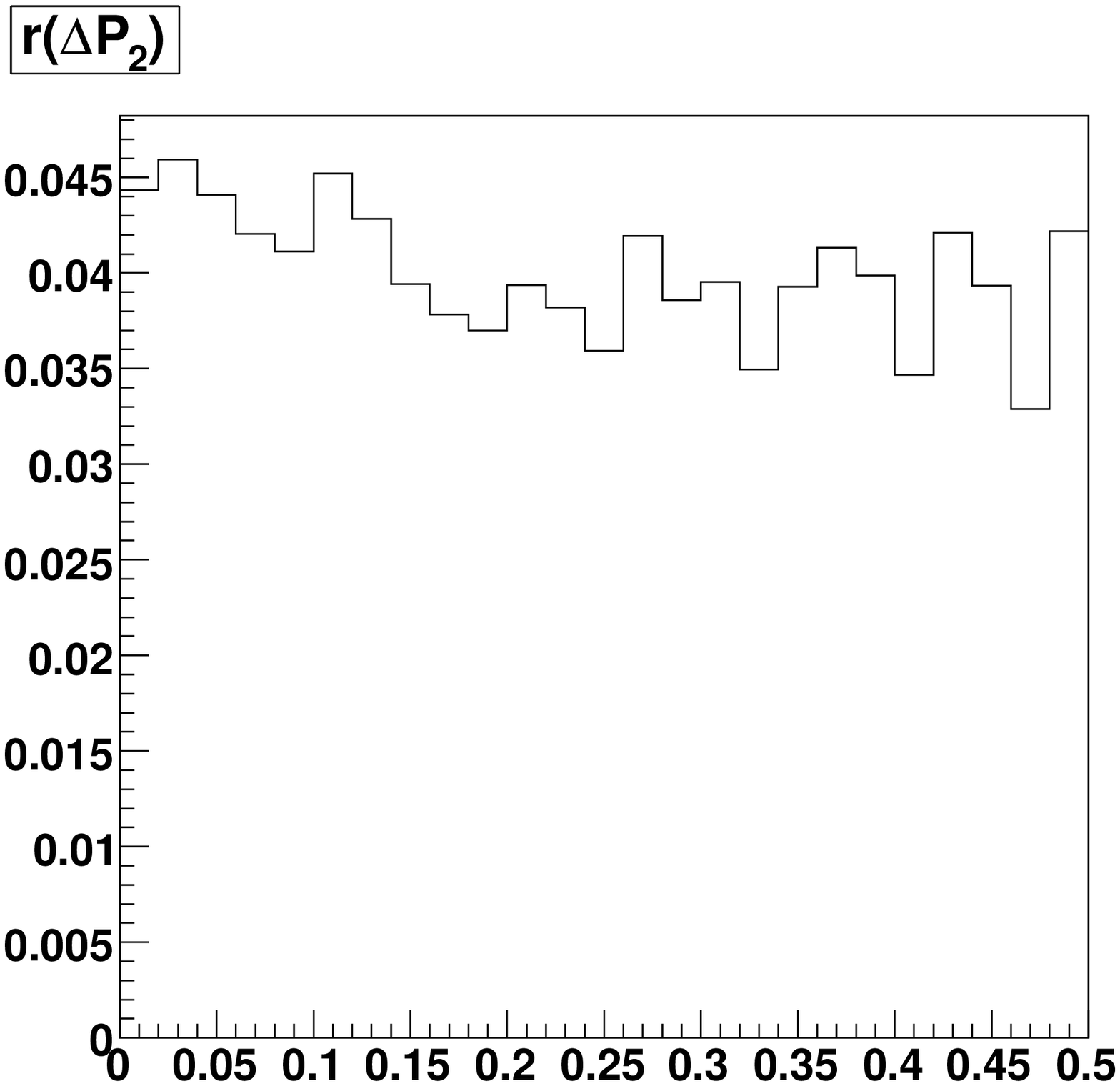}
 \includegraphics[width=.32\textwidth]{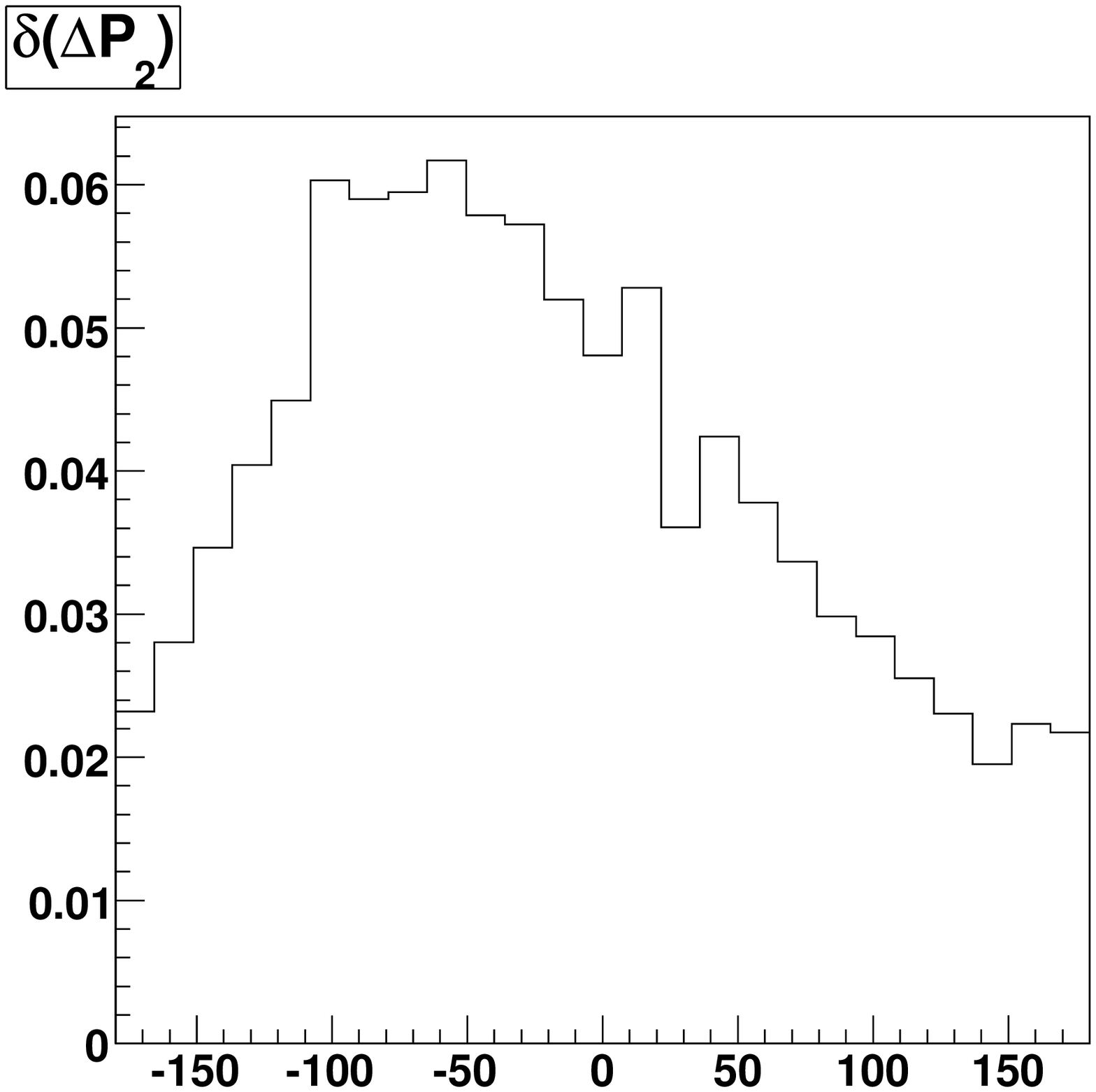}
 \includegraphics[width=.32\textwidth]{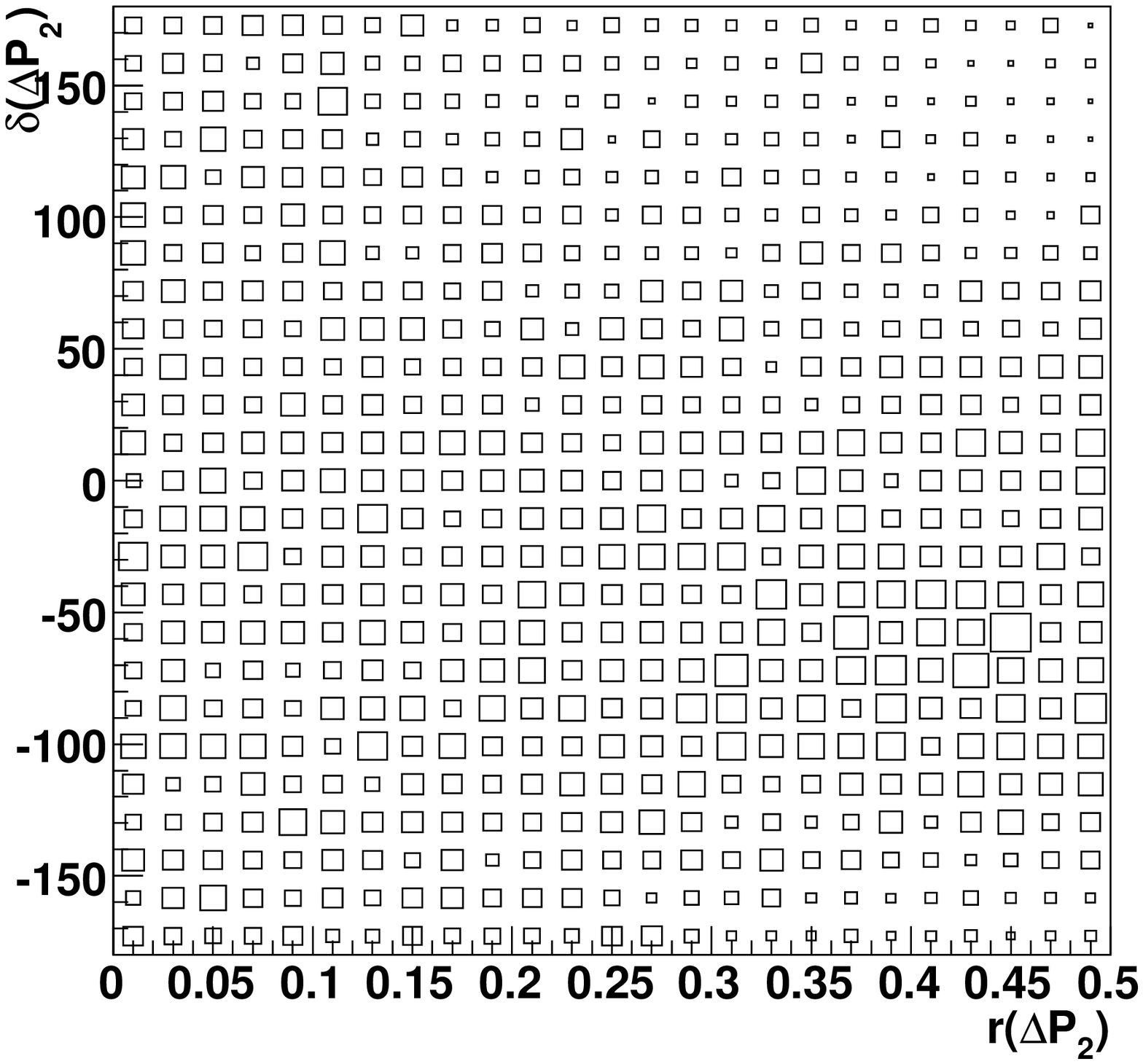}
 \end{center}
\caption{1D and 2D p.d.f.'s obtained from the global fit for the
  parameters $r(\Delta P_1)$, $\delta(\Delta P_1)$ and $r(\Delta
  P_2)$, $\delta(\Delta P_2)$ defined in
  eqs.~(\ref{eq:params1}).\label{fig:had2}} 
\end{figure*}

Our procedure is to fit the hadronic parameters to the experimental
data, taking into account the hierarchy between leading and subleading
terms in the $1/m_b$ expansion by imposing an upper bound to
subleading corrections. Only the correction to the dominant penguin
amplitude is well determined by the fit. The information on the
subdominant terms is limited, while their presence contributes to the
theoretical uncertainty. The theoretical error on the predicted
observables is thus determined by the allowed range for the subleading
parameters. While quantifying this range is somewhat arbitrary,
extreme situations in which the leading and subleading terms are
comparable would imply a failure of the infinite mass limit. Of
course, one has to be careful about possible parametric or dynamical
enhancements which could invalidate the power counting.
Chirally-enhanced terms in $B \to K \pi$ amplitudes are well-known
examples of terms that are formally subleading but numerically of
$\mathcal{O}(1)$. We have therefore included them in the leading
factorized amplitudes. We now quantify the allowed ranges we use for
subleading corrections. To this aim, we write each parameter as
follows:
\begin{eqnarray}
  E_1 &=& E_1^\mathrm{F} + F\,r(E_1)\,,\nonumber \\
  E_2 &=& E_2^\mathrm{F} + F\,r(E_2) e^{i\delta(E_2)}\,,\nonumber \\
  A  &=& A^\mathrm{F} + F\,r(A) e^{i\delta(A)}\,,       \nonumber \\
  P   &=& P^\mathrm{F} + F\, r(P)e^{i \delta(P)}\,,      \nonumber \\
  \Delta P_1 &=& \Delta P_1^\mathrm{F}+F\,\alpha_\mathrm{em}\, r(\Delta P_1)e^{i \delta(\Delta P_1)}\,,  \nonumber \\
  \Delta P_2 &=& \Delta P_2^\mathrm{F}+F\,\alpha_\mathrm{em}\, r(\Delta P_2)e^{i \delta(\Delta P_2)}\,,
  \label{eq:params1}
\end{eqnarray}
where the factorized amplitudes in the limit $m_b \to \infty$ are
\begin{eqnarray}
  E_1^\mathrm{F} &=& A_{\pi K}\biggl(-\alpha_1-\alpha_4^u+\alpha_4^c-\alpha_{4,EW}^u+\alpha_{4,EW}^c\biggr)\,, \nonumber \\
  E_2^\mathrm{F} &=&
  A_{K\pi}\biggl(-\alpha_2-\frac{3}{2}(\alpha_{3,EW}^u-\alpha_{3,EW}^c)\biggr)\nonumber
  \\ &&+A_{\pi K}\biggl(\alpha_{4}^u-\alpha_{4}^c-\frac{1}{2}(\alpha_{4,EW}^u-\alpha_{4,EW}^c)\biggr)\,,  \nonumber \\
  A^\mathrm{F} &=&   A_{\pi K}\biggl( -\alpha_{4}^u+\alpha_{4}^c +\frac{1}{2}
  (\alpha_{4,EW}^u-\alpha_{4,EW}^c )\biggr)\,, \nonumber\\
  P^\mathrm{F}   &=& A_{\pi K}\biggl(-\alpha_4^c+\frac{1}{2}\alpha_{4,EW}^c\biggr)\,,  \nonumber \\
  \Delta P_1^\mathrm{F} &=& - A_{\pi K} \frac{3}{2}\alpha_{4,EW}^c\,,  \nonumber \\
  \Delta P_2^\mathrm{F} &=& - A_{K \pi} \frac{3}{2}\alpha_{3,EW}^c\,,
  \label{eq:fact}
\end{eqnarray}
in terms of the parameters $\alpha$ defined in eq.~(31) of
ref.~\cite{bn}. We note that we have discarded non-factorizable contributions
to the chirally enhanced terms.
Furthermore, 
\begin{eqnarray}
  A_{\pi K}&=& G_F/\sqrt{2}m_B^2f_K F_\pi(0)\,, \nonumber \\
  A_{K \pi}&=& G_F/\sqrt{2}m_B^2f_\pi F_K(0)\,.
  \label{eq:afact}
\end{eqnarray}
The coefficient $F$ in eqs.~(\ref{eq:params1}) sets the normalization
of subleading corrections and is equal to $A_{\pi K}$ computed using
the central value of the form factor. The phase convention is chosen
such that the power correction to $E_1$ is real.

The subleading terms in units of $F$ are given by $r(X)=[0,0.5]$ for
$X=\{E_1,E_2,A,\Delta P_1,\Delta P_2\}$. Since $r(P)$ is very well
determined by the fit, for computational efficiency we used
$r(P)=[0,0.2]$. For the sake of comparison, Ref.~\cite{bn} quotes a
value of $0.09^{+0.32}_{-0.09}$ for the contribution to $r(P)$ from
penguin annihilation, compatible with the range we use. All strong
phases vary in the range $[-\pi,\pi]$.

\begin{table}
\begin{center}
\begin{tabular}{|c|c|c|c|}
\hline
$f_\pi$        & $0.1307$~GeV              & $f_K$                       & $0.1598$~GeV             \\ 
$F^{B \to \pi}$ & $0.27 \pm 0.08$          & $F^{B \to K}/F^{B \to \pi}$ & $1.20 \pm 0.10$          \\
$\tau_{B^0}$    & $1.546\cdot 10^{-12}$ ps & $\tau_{B^+}$                & $1.674\cdot 10^{-12}$ ps \\
$m_B$          & $5.2794$~GeV      & $f_{B}$                   & $0.189 \pm 0.027$~GeV    \\
$m_\pi$        & $0.14$~GeV         &$m_K$                      & $0.493677$~GeV     \\
\hline
$\lambda$ & $0.2258\pm 0.0014$ & $A$ & $0.810\pm 0.011$ \\
$\bar\rho$ & $0.154\pm 0.022$ & $\bar\eta$ & $0.342\pm 0.014$ \\
\hline
\end{tabular}
\end{center}
\caption{Input values used in the analysis. Form factors are taken
  from lattice QCD calculations~\cite{latticeQCD}. CKM parameters have been
  taken from ref.~\cite{hep-ph/0606167}. Wave function parameters
  can be found in Table~1 of ref.~\cite{bn}.}
\label{tab:charminginput}
\end{table}

Using the ranges above for the hadronic parameters and the input
parameters reported in Table~\ref{tab:charminginput}, we perform a fit
to the data in Table~\ref{tab:res} using the method described in
ref.~\cite{utfit2000}. Flat priors are used for the hadronic parameters.
Two sets of results are summarized in
Table~\ref{tab:res}.  On one hand, when using all the experimental
information as input we test the consistency of the SM description of
the decay amplitudes in a {\it global fit}. On the other hand, by
removing one of the inputs from the fit we obtain a {\it prediction}
of the corresponding experimental observable, using all the other
inputs to constrain the hadronic parameters.

Two main results are obtained from the {\it global fit}: {\it i})
the BR values are well reproduced, and they are fairly insensitive to
the $1/m_b$ contributions, but for the CKM-enhanced
charming penguin $P$. {\it ii}) The values of the $A_\mathrm{CP}$ are well
reproduced, thanks to the $1/m_b$ contributions. In
particular, the presence of $\Delta P_2$ ($E_2+A$) in the CKM-enhanced
(CKM-suppressed) part of the $B^+ \to K^+\pi^0$ amplitude (see
Eq.~(\ref{eq:ampli})) allows to obtain simultaneously a positive value
of $A_\mathrm{CP}(K^+ \pi^0)$ and a negative value of
$A_\mathrm{CP}(K^+ \pi^-)$.  This is shown in the left plot
of Fig.~\ref{fig:acpdiff_sk0pi0}, where the output distribution of
$\Delta A_\mathrm{CP}$ is fully consistent with the experimental world
average $\Delta A_\mathrm{CP} = 0.148 \pm 0.028$.  

The results for the hadronic parameters are shown in
Figs.~\ref{fig:had0}--\ref{fig:had2}. Both the charming penguin
parameters $r(P)$ and $\delta(P)$ are well determined, in agreement
with the old results of ref.~\cite{charming}.  In particular, $r(P)$
is found to be of ${\cal O}(1/m_b)$, as expected from the power
expansion in QCD factorization. Small values of $r(A)$ are favoured,
although values as large as $0.5$ are not excluded. However, a large
$r(A)$ requires a $\delta(A)$ small and negative.  The corrections to
the parameter $E_2$, on the other hand, are pushed towards the upper
half of the allowed range, namely $0.3\div0.5$, showing a preference
for a large correction to the color-suppressed emission
amplitude~\cite{puzzle2,largee2,np2}.  However, we have checked that
the p.d.f.\ for $r(E_2)$ falls for values larger than $0.6$ (although
there are other allowed regions for $r(E_2) \gg 1$, see below).  No
information on the other parameters can be extracted from the fit, but
for a slight modulation of the phases in the region of absolute values
close to the upper bound. 

\begin{figure}[tb]
 \begin{center}
 \includegraphics[width=.35\textwidth]{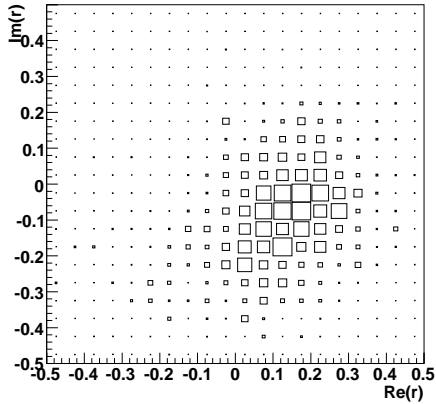}
\end{center}
\caption{P.d.f. obtained from the global  fit for Im$(r)$
  vs. Re$(r)$.}
\label{fig:rsu3}
\end{figure}

We have checked that the result for $(\Delta P_1+ \Delta
P_2)/(E_1+E_2)$ is in agreement with the prediction of
ref.~\cite{minchiatesu3ew} (obtained in the SU(3) limit neglecting
left-right electroweak penguins). To quantify this statement, we
define, following ref.~\cite{kstarpi}, the SU(3) breaking ratio of
matrix elements
\begin{equation}
  \label{eq:rsu3}
  r=\frac{\langle K\pi (I=3/2)| Q_-|B\rangle}{\langle K\pi (I=3/2)|
    Q_+|B\rangle}\,. 
\end{equation}
In factorization, this ratio is tiny due to the fact that $f_k F^{B
  \to \pi}\sim f_\pi F^{B \to K}$, so that $r \sim
\left\vert\frac{f_{K} F^{B\to \pi}-f_\pi F^{B\to K}} {f_{K} F^{B\to
      \pi}+f_\pi F^{B\to K}}\right\vert \sim O(10^{-2})$. However,
this cancellation is not related to SU(3) (in fact, it also holds for
$B \to K^* \pi$, where the SU(3) argument does not apply). More
generally, one expects $\vert r\vert \lesssim 20\%$. In
Fig.~\ref{fig:rsu3} we present the value of $r$ obtained from our
global fit, yielding $\vert r \vert = 0.20 \pm 0.08$. The fit is fully
compatible with the general expectations on SU(3) breaking. The
factorization predictions are also compatible with the fit result,
although the fit prefers larger values of SU(3) breaking.

Going back to the parameters on the r.h.s. of Eq.~(\ref{eq:params1}),
we can conclude that $P_1^\mathrm{GIM}$ is not the dominant source of
power corrections in $E_1$, $E_2$ and $A$ as this would imply definite
correlations among $E_1$, $E_2$ and $A$ which are not observed.

Another mechanism for reproducing the $K\pi$ data proposed in the
literature~\cite{np,bf,np2} is a NP contribution enhancing the EWP
amplitudes with a new CP-violating phase. While we do not include NP
phases in our analysis, we checked that removing subleading
corrections to emissions and annihilations and allowing $r(\Delta
P_2)$ to violate the $1/m_b$ power counting, it is not possible to
reproduce the $K\pi$ data.

\begin{figure}[tb]
 \begin{center}
 \includegraphics[width=.38\textwidth]{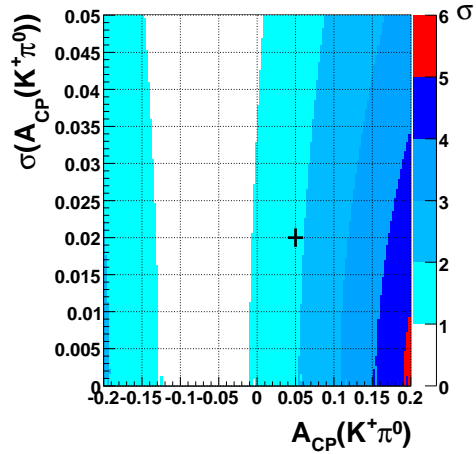}
 \includegraphics[width=.38\textwidth]{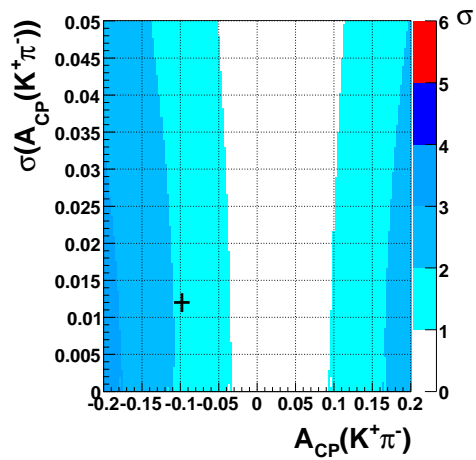}
 \end{center}
\caption{Compatibility plots for $A_\mathrm{CP}(K^+ \pi^0)$ (upper) and
$A_\mathrm{CP}(K^+ \pi^-)$ (lower). The cross denotes the experimental
values. The colour code indicates the level of compatibility with
the SM prediction.\label{fig:compat}}
\end{figure}

The {\it predictions} for the BR, obtained by removing them one by one
from the fit, show that the observed values can be easily explained,
all the values being in the $\pm 1\sigma$ range, the error on the
prediction being comparable to the experimental one. 
On the other hand, the error on the predictions for $A_\mathrm{CP}$ is much
larger than the experimental precision (up to a factor six for
$A_\mathrm{CP}(K^+ \pi^-)$). Within these large uncertainties,
the predictions are in agreement with the experimental values at the
1--2$\sigma$ level, as shown in Fig.~\ref{fig:compat}.

\begin{figure*}[tb]
  \centering
  \includegraphics[width=.32\textwidth]{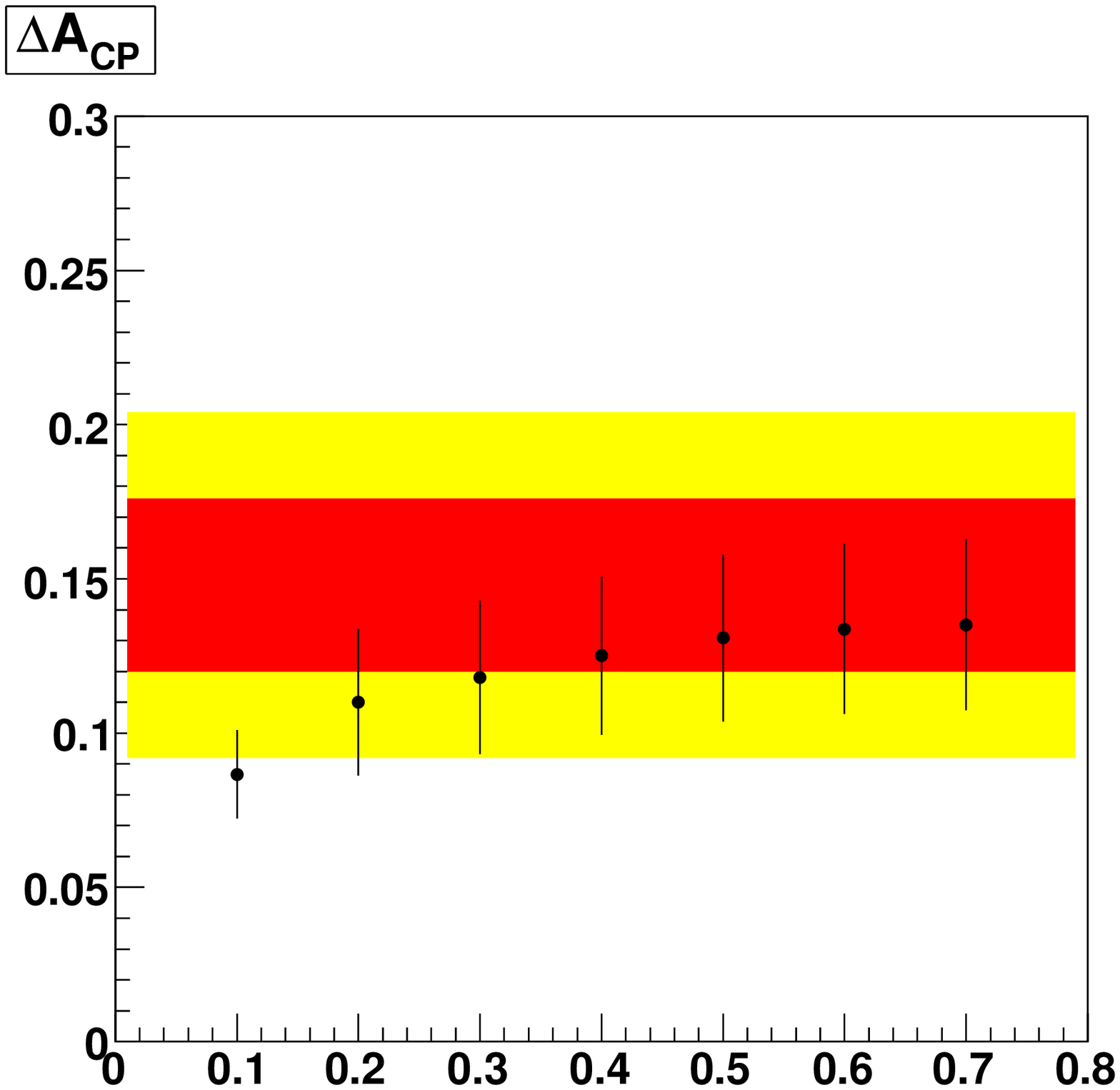}  
  \includegraphics[width=.32\textwidth]{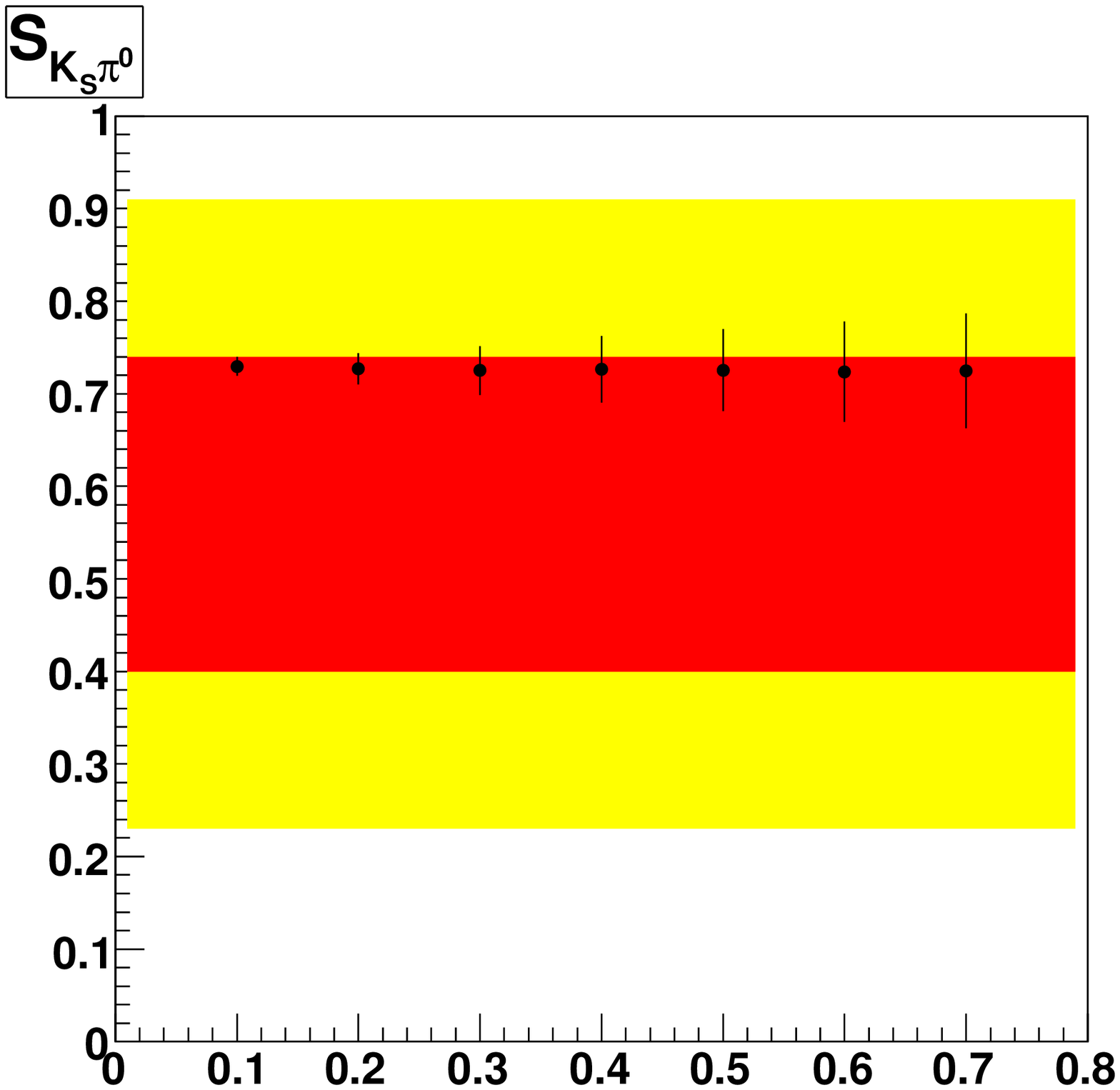}  
  \includegraphics[width=.32\textwidth]{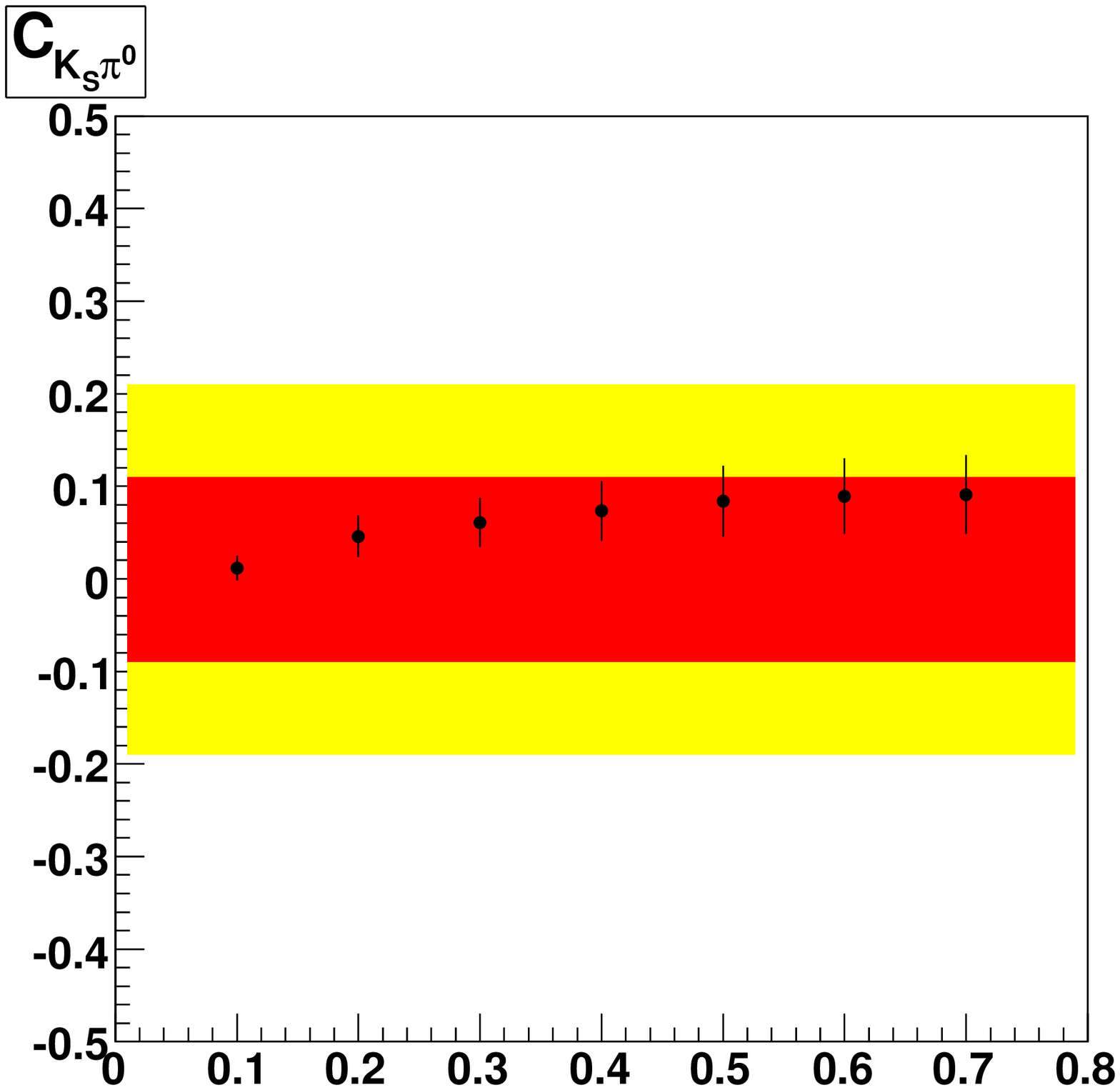}  
  \caption{Some fit results as functions of the upper bound on power corrections.}
  \label{fig:ul}
\end{figure*}

The choice of the upper limit for the subleading terms used in our fit
clearly dictates the theoretical error associated to the fit
predictions. For example, raising the upper limit from 0.5 to 1 the
error on the fit prediction for $\Delta A_\mathrm{CP}$ increases from
$0.06$ to $0.09$. On the other hand, the results of the global fit are
fairly independent of this choice provided that the upper limit is
large enough, as shown in Fig.~\ref{fig:ul}. In fact, our point is
that a good fit of the experimental data can be obtained for
subleading terms compatible with power counting. Once a good fit is
obtained, the dependence on the upper bound becomes negligible. On the
other hand too small values of the upper limit would result in a worse
agreement between the theory and the data, showing that the
factorization formulae need to be completed with non-perturbative
$1/m_b$ corrections to give a good description of the data.

\begin{figure}[tb]
 \begin{center}
 \includegraphics[width=.38\textwidth]{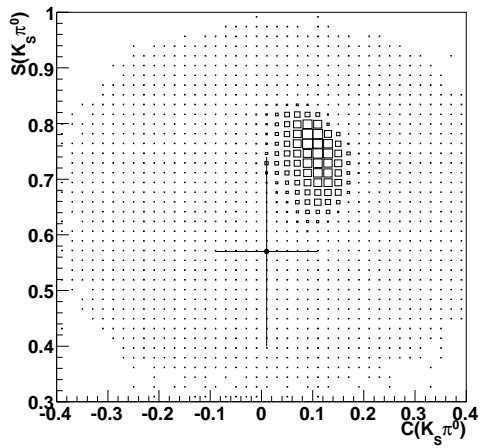}
 \end{center}
\caption{2D p.d.f. in the $S_{K_S\pi^0}$--$C_{K_S\pi^0}$ plane obtained without
using the experimental information on both observables. The cross
represents the experimental result.\label{fig:SvsC}}
\end{figure}

Removing both $S_{K_S\pi^0}^\mathrm{exp}$ and
$C_{K_S\pi^0}^\mathrm{exp}$ from the fit, an interesting prediction
can still be obtained for the parameters of the $B^0 \to K_S \pi^0$
time-dependent CP asymmetry. We get $C_{K_S\pi^0}=0.10\pm 0.04$, in
good agreement with the experimental measurement, and
$S_{K_S\pi^0}=0.74\pm 0.04$, which is compatible with the experimental
world average at the $\sim 1\sigma$ level.  In Fig.~\ref{fig:SvsC} we
show the selected region on the $S_{K_S\pi^0}$--$C_{K_S\pi^0}$ plane,
compared to the
experimental determination.  Both the prediction and the measurement
are limited by the experimental precision, since the other $K\pi$ data
are a crucial ingredient in our fit. For example, reducing all
experimental errors by a factor of two, the error on the fit
prediction for $S_{K_S\pi^0}$ decreases to $0.03$, while the error on
$C_{K_S\pi^0}$ decreases to $0.02$. It is then mandatory to improve
the experimental information. Considering the difficulties related to the
study of $B^0 \to K^0 \pi^0$ in the crowded environment of LHC,
SuperB~\cite{superB} appears as the ideal facility to accomplish this
task.

Recently, ref.~\cite{arXiv:0806.2900} pointed out a correlation
between $S_{K_S\pi^0}$ and $C_{K_S\pi^0}$. Using the experimental
value of $C_{K_S\pi^0}$ they obtained
$S_{K_S\pi^0}=0.99^{+0.01+0.00+0.00}_{-0.07-0.10-0.06}$.  Similar
results were found in ref.~\cite{arXiv:0807.3080}. Both papers make
the following assumptions: $\Delta I=3/2$ amplitude fixed from
$\pi\pi$ data using $SU(3)$ symmetry (neglecting also left-right
electroweak penguins). Under these assumptions, they solve for the
amplitudes $A_{00}=A(B^0\to K^0\pi^0)$, $A_{+-}=A(B^0\to K^+\pi^-)$
and the CP-conjugate ones $\bar A_{00,+-}$, up to a four-fold
ambiguity.  This ambiguity can be lifted using phenomenological
arguments partly based on $SU(3)$ and involving charged $B\to K \pi$
modes, further neglecting annihilations.  Both papers find a large
value of $\phi_{00}=\mathrm{arg}(A_{00} \bar A^*_{00})\sim 42^\circ$
leading to a prediction for $S_{K_S\pi^0}$ close to one. We have
repeated the analysis and were able to reproduce the results of
refs.~\cite{arXiv:0806.2900,arXiv:0807.3080}. In addition, we computed
the values of the relevant hadronic parameters (in our notation:
$E_{1,2}$ and $P$) corresponding to the four solutions for the
amplitudes $A_{00}$,$A_{+-}$,$\bar A_{00}$,$\bar A_{+-}$. In
particular, neglecting annihilations and $\Delta I=1/2$ EWP, the
solution with $\phi_{00}\sim 42^\circ$ has a value of $P$ giving a
BR$(B^+\to K^0\pi^+)\sim 18\times 10^{-6}$, incompatible with the
measured value $(23.1\pm 1.0)\times 10^{-6}$. In any case, one gets a
huge value of $\vert E_2/E_1\vert$: using the input of
ref.~\cite{arXiv:0807.3080}, we find $E_2/E_1=1.9 e^{-i 176^\circ}$.
Clearly this value is not compatible with factorization and would
imply a breakdown of the heavy quark expansion. In our fit, by
limiting the range of the power corrections, we discarded this
possibility. In fact, we have shown that a good agreement with the
experimental data is possible without introducing huge corrections to
factorization. Another recent analysis, presented in
ref.~\cite{arXiv:0803.3729}, obtained a good agreement with
experimental data, fixing the ratio of EWP to current-current operator
matrix elements using QCD factorization and fitting all other matrix
elements. The range found in ref.~\cite{arXiv:0803.3729} for $\vert
E_2/E_1\vert=[0.52,3]$ can possibly be compatible with both our
findings and the results of
refs.~\cite{arXiv:0806.2900,arXiv:0807.3080}.  Indeed it may overlap
with our findings in the lower range allowed for $\vert E_2/E_1 \vert$
but also with those of refs.~\cite{arXiv:0806.2900,arXiv:0807.3080} in
the upper range where $\vert E_2/E_1\vert$ violates the $1/m_b$ power
counting.

In this letter, we presented a data-driven method to estimate the
hadronic uncertainties in $B\to K\pi$ amplitudes
compatible with the $1/m_b$ expansion.  This is a basic requirement to
meaningfully look for NP in these channels.  We found that $K\pi$ data
can be accounted for by the SM, including direct CP violation. CP
violating asymmetries are predicted with a large uncertainty, except
for  $S_{K_S\pi^0}$ and $C_{K_S\pi^0}$, where the theoretical error is
much smaller than the experimental one. Thus, these asymmetries are a
better place to look for NP than direct CP violation in the other $B
\to K \pi$ decay modes, where possible NP contributions are obscured
by hadronic uncertainties.\\
~\\
\noindent\textbf{Note added}\\
During the completion of this work, we were informed that similar
results have been obtained by M.~Duraisamy and A.~Kagan in an ongoing
analysis of power corrections to $B\to PP$, $PV$, and $VV$ decays.
Earlier results by the same group can be found in ref.~\cite{kaganCKM06}.\\

We thank R.~Fleischer for useful discussions.
We acknowledge partial support from RTN European contracts
MRTN-CT-2006-035482 ``FLAVIAnet'' and MRTN-CT-2006-035505
``Heptools''. M.C.{} is associated to the Dipartimento di Fisica,
Universit\`a di Roma Tre. E.F.{} and L.S.{} are associated to the
Dipartimento di Fisica, Universit\`a di Roma ``La Sapienza''.

\end{document}